# STABILITY OF THE RELATIVE EQUILIBRIA OF A RIGID BODY IN A $J_2$ GRAVITY FIELD

Yue Wang,[*] Haichao Gui[†], and Shijie Xu[‡]

The motion of a point mass in the $J_2$ problem is generalized to that of a rigid body in a $J_2$ gravity field. Different with the original $J_2$ problem, the gravitational orbit-rotation coupling of the rigid body is considered in this generalized problem. The linear stability of the classical type of relative equilibria of the rigid body, which have been obtained in our previous paper, is studied in the framework of geometric mechanics with the second-order gravitational potential. Non-canonical Hamiltonian structure of the problem, i.e., Poisson tensor, Casimir functions and equations of motion, are obtained through a Poisson reduction process by means of the symmetry of the problem. The linear system matrix at the relative equilibria is given through the multiplication of the Poisson tensor and Hessian matrix of the variational Lagrangian. Based on the characteristic equation of the linear system matrix, the conditions of linear stability of the relative equilibria are obtained. With the stability conditions obtained, the linear stability of the relative equilibria is investigated in details in a wide range of the parameters of the gravity field and the rigid body. We find that both the zonal harmonic $J_2$ and the characteristic dimension of the rigid body have significant effects on the linear stability. Similar to the attitude stability in a central gravity field, the linear stability region is also consisted of two regions that are analogues of the Lagrange region and the DeBra-Delp region. Our results are very useful for the studies on the motion of natural satellites in our solar system.

## INTRODUCTION

The $J_2$ problem, also called the main problem of artificial satellite theory, in which the motion of a point mass in a gravity field truncated on the zonal harmonic $J_2$ is studied, is an important problem in the celestial mechanics and astrodynamics[1]. The $J_2$ problem has its wide applications in the orbital dynamics and orbital design of spacecraft. This classical problem has been studied by many authors, such as Reference [1] and the literatures cited therein.

However, neither natural nor artificial celestial bodies are point masses or have spherical mass distribution. One of the generalizations of the point mass model is the rigid body model. Because of the non-spherical mass distribution, the orbital and rotational motions of the rigid body are coupled through the gravity field. The orbit-rotation coupling may cause qualitative effects on the motion, which are more significant when the ratio of the dimension of rigid body to the orbit radius is larger. The orbit-rotation coupling and its qualitative effects have been discussed in several works on the motion of a rigid body or gyrostat in a central gravity field[2]–[5]. In Reference [6], the orbit-rotation coupling of a rigid satellite around a spheroid planet was assessed. It was found that the significant orbit-rotation coupling should be considered for a spacecraft orbiting a small asteroid or an irregular natural satellite orbiting a planet.

[*] PhD Candidate, Department of Aerospace Engineering, School of Astronautics, Beihang University, Beijing, 100191, China. e-mail: ywang@sa.buaa.edu.cn
[†] PhD Candidate, Department of Aerospace Engineering, School of Astronautics, Beihang University, Beijing, 100191, China. e-mail: guihaichao@gmail.com
[‡] Professor, Department of Aerospace Engineering, School of Astronautics, Beihang University, Beijing, 100191, China. e-mail: starsjxu@yahoo.com.cn

The effects of the orbit-rotation coupling have also been considered in many works on the Full Two Body Problem (F2BP), in which the rotational and orbital motions of two rigid bodies interacting through their mutual gravitational potential are studied. A sphere-restricted model of F2BP, in which one body is assumed to be a homogeneous sphere, has been studied broadly by many scholars, such as Kinoshita[7], Barkin[8], Aboelnaga and Barkin[9], Beletskii and Ponomareva[10], Scheeres[11], Breiter et al.[12], Balsas et al.[13], Bellerose and Scheeres[14], and Vereshchagin et al.[15]. There are also several works on the more general models of F2BP, in which both bodies are non-spherical, such as the works by Maciejewski[16], Scheeres[17][18], Koon et al.[19], Boué and Laskar[20], McMahon and Scheeres[21].

When the dimension of the rigid body is very small in comparison with the orbital radius, the orbit-rotation coupling is not significant. In the case of an artificial Earth satellite, the point mass model of the $J_2$ problem works very well. However, when a spacecraft orbiting an asteroid or an irregular natural satellite orbiting a planet, such as Phobos, is considered, the mass distribution of the considered body is far from a sphere and the dimension of the body is not small anymore in comparison with the orbital radius. In these cases, the orbit-rotation coupling causes significant effects and should be taken into account in the precise theories of the motion, as shown by Koon et al.[19], Scheeres[22], Wang and Xu[6]. For the high-precision applications in the motions of a spacecraft orbiting a spheroid asteroid, or an irregular natural satellite orbiting a dwarf planet or planet, we have generalized the $J_2$ problem to the motion of a rigid body in a $J_2$ gravity field in our previous paper[23]. In that paper, the relative equilibria of the rigid body were determined from a global point of view in the framework of geometric mechanics.

Through the non-canonical Hamiltonian structure of the problem, the geometric mechanics provides a systemic and effective method for determining the stability of the relative equilibria, as shown by Beck and Hall[24]. The linear stability of the classical type of relative equilibria, which have been already obtained in Reference [23], will be studied further in this paper with the geometric mechanics. Through the stability properties of the relative equilibria, it is sufficient to understand the dynamical properties of the system near the relative equilibria to a big extent.

The equilibrium configuration exists generally among the natural celestial bodies in our solar system. It is well known that many natural satellites of big planets evolved tidally to the state of synchronous motion[25]. Notice that the gravity field of the big planets can be well approximated by a $J_2$ gravity field. The results on the stability of the relative equilibria in our problem are very useful for the studies on the motion of many natural satellites. We also make comparisons with previous results on the stability of the relative equilibria of a rigid body in a central gravity field, such as References [2] and [5]. The influence of the zonal harmonic $J_2$ on the stability of the relative equilibria is discussed in details.

**NON-CANONICAL HAMILTONIAN STRUCTURE AND RELATIVE EQUILIBRIA**

The problem we studied here is the same as in Reference [23]. As described in Figure 1, we consider a small rigid body *B* in the gravity field of a massive axis-symmetrical body *P*. Assume that *P* is rotating uniformly around its axis of symmetry, and the mass center of *P* is stationary in the inertial space. The gravity field of *P* is approximated through truncation on the second zonal harmonic $J_2$. The inertial reference frame is defined as $S=\{e_1, e_2, e_3\}$ with its origin *O* attached to the mass center of *P*. $e_3$ is along the axis of symmetry of *P*. The body-fixed reference frame of the rigid body is defined as $S_b=\{i, j, k\}$ with its origin *C* attached to the mass center of *B*. The frame $S_b$ coincides with the principal axes reference frame of the rigid body *B*.

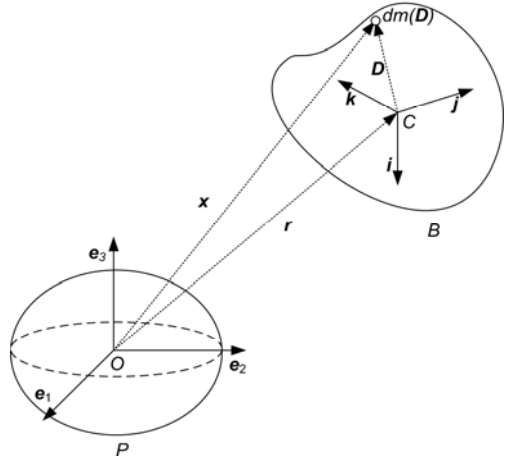

**Figure 1. A small rigid body *B* in the $J_2$ gravity field of a massive axis-symmetrical body *P*.**

In Reference [23], a Poisson reduction was applied on the original system by means of the symmetry of the problem. After the reduction process, the non-canonical Hamiltonian structure, i.e., Poisson tensor, Casimir functions and equations of motion, and a classical kind of relative equilibria of the problem were obtained. Here we only give the basic description of the problem and list the main results obtained by us there, see that paper for the details. The attitude matrix of the rigid body *B* with respect to the inertial frame *S* is denoted by *A*,

$$A = [i, j, k] \in SO(3), \qquad (1)$$

where the vectors *i*, *j* and *k* are expressed in the frame *S*, and *SO*(3) is the 3-dimensional special orthogonal group. *A* is the coordinate transformation matrix from the frame $S_b$ to the frame *S*. If $W = [W^x, W^y, W^z]^T$ a vector expressed in frame $S_b$, its components in frame *S* can be given by

$$w = AW. \qquad (2)$$

We define *r* as the position vector of point *C* with respect to *O* in frame *S*. The position vector of a mass element *dm(D)* of the body *B* with respect to *C* in frame $S_b$ is denoted by *D*, then the position vector of *dm(D)* with respect to *O* in frame *S*, denoted by *x*, is

$$x = r + AD. \qquad (3)$$

Therefore, the configuration space of the problem is the Lie group

$$Q = SE(3), \qquad (4)$$

known as the special Euclidean group of three space with elements (*A*, *r*) that is the semidirect product of *SO*(3) and $\mathbb{R}^3$. The elements *Ξ* of the phase space, the cotangent bundle $T^*Q$, can be written in the following coordinates

$$\Xi = (A, r; A\hat{\Pi}, p), \qquad (5)$$

where $\boldsymbol{\Pi}$ is the angular momentum expressed in the body-fixed frame $S_b$ and $\boldsymbol{p}$ is the linear momentum of the rigid body expressed in the inertial frame $S$[26]. The hat map $\wedge: \mathbb{R}^3 \to so(3)$ is the usual Lie algebra isomorphism, where $so(3)$ is the Lie Algebras of Lie group $SO(3)$.

The phase space $T^*Q$ carries a natural symplectic structure $\omega = \omega^{SE(3)}$, and the canonical bracket associated to $\omega$ can be written in coordinates $\boldsymbol{\Xi}$ as

$$\{f, g\}_{T^*Q}(\boldsymbol{\Xi}) = \langle D_A f, D_{A\hat{\boldsymbol{\Pi}}} g \rangle - \langle D_A g, D_{A\hat{\boldsymbol{\Pi}}} f \rangle + \left(\frac{\partial f}{\partial \boldsymbol{r}}\right)^T \frac{\partial g}{\partial \boldsymbol{p}} - \left(\frac{\partial g}{\partial \boldsymbol{r}}\right)^T \frac{\partial f}{\partial \boldsymbol{p}}, \tag{6}$$

for any $f, g \in C^\infty(T^*Q)$, $\langle \cdot, \cdot \rangle$ is the pairing between $T^*SO(3)$ and $TSO(3)$, and $D_B f$ is a matrix whose elements are the partial derivates of the function $f$ with respect to the elements of matrix $\boldsymbol{B}$ respectively[26].

The Hamiltonian of the problem $H: T^*Q \to \mathbb{R}$ is given as follows

$$H = \frac{|\boldsymbol{p}|^2}{2m} + \frac{1}{2}\boldsymbol{\Pi}^T \boldsymbol{I}^{-1} \boldsymbol{\Pi} + V \circ \tau_{T^*Q}, \tag{7}$$

where $m$ is the mass of the rigid body, the matrix $\boldsymbol{I} = diag\{I_{xx}, I_{yy}, I_{zz}\}$ is the tensor of inertia of the rigid body and $\tau_{T^*Q}: T^*Q \to Q$ is the canonical projection.

According to Reference [23], the gravitational potential $V: Q \to \mathbb{R}$ up to the second order is given in terms of moments of inertia as follows:

$$V = V^{(0)} + V^{(2)} = -\frac{GM_1 m}{R} - \frac{GM_1}{2R^3}\left[tr(\boldsymbol{I}) - 3\bar{\boldsymbol{R}}^T \boldsymbol{I} \bar{\boldsymbol{R}} + \varepsilon m - 3\varepsilon m(\boldsymbol{\gamma} \cdot \bar{\boldsymbol{R}})^2\right], \tag{8}$$

where $G$ is the Gravitational Constant, and $M_1$ is the mass of the body $P$. The parameter $\varepsilon$ is defined as $\varepsilon = J_2 a_E^2$, where $a_E$ is the mean equatorial radius of $P$. $\boldsymbol{\gamma}$ is the unit vector $\boldsymbol{e}_3$ expressed in the frame $S_b$. $\boldsymbol{R} = A^T \boldsymbol{r}$ is the position vector of the mass center of $B$ expressed in frame $S_b$. Note that $R = |\boldsymbol{R}|$ and $\bar{\boldsymbol{R}} = \boldsymbol{R}/R$.

The $J_2$ gravity field is axis-symmetrical with axis of symmetry $\boldsymbol{e}_3$. According to Reference [26], the Hamiltonian of the system is $S^1$-invariant, namely the system has symmetry, where $S^1$ is the one-sphere. Using this symmetry, we have carried out a reduction, induced a Hamiltonian on the quotient $T^*Q/S^1$, and expressed the dynamics in terms of appropriate reduced variables in Reference [26], where $T^*Q/S^1$ is the quotient of the phase space $T^*Q$ with respect to the action of $S^1$. The reduced variables in $T^*Q/S^1$ can be chosen as

$$\boldsymbol{z} = \left[\boldsymbol{\Pi}^T, \boldsymbol{\gamma}^T, \boldsymbol{R}^T, \boldsymbol{P}^T\right]^T \in \mathbb{R}^{12}, \tag{9}$$

where $\boldsymbol{P} = A^T \boldsymbol{p}$ is the linear momentum of the body $B$ expressed in the body-fixed frame $S_b$[26]. The projection from $T^*Q$ to $T^*Q/S^1$ is given by

$$\Psi(A, \boldsymbol{r}; A\hat{\boldsymbol{\Pi}}, \boldsymbol{p}) = \left[\boldsymbol{\Pi}^T, \boldsymbol{\gamma}^T, \boldsymbol{R}^T, \boldsymbol{P}^T\right]^T. \tag{10}$$

There is a unique non-canonical Hamiltonian structure on $T^*Q/S^1$ such that $\Psi$ is a Poisson map. That is to say, there is a unique Poisson bracket $\{\cdot, \cdot\}_{\mathbb{R}^{12}}(\boldsymbol{z})$ satisfying[27]

$$\{f, g\}_{\mathbb{R}^{12}}(z) \circ \Psi = \{f \circ \Psi, g \circ \Psi\}_{T^*Q}(\Xi), \tag{11}$$

for any $f, g \in C^\infty(\mathbb{R}^{12})$, where $\{\cdot, \cdot\}_{T^*Q}(\Xi)$ is the natural canonical bracket given by Eq. (6).

According to Reference [26], the Poisson bracket $\{\cdot, \cdot\}_{\mathbb{R}^{12}}(z)$ can be written as follows:

$$\{f, g\}_{\mathbb{R}^{12}}(z) = (\nabla_z f)^T \boldsymbol{B}(z)(\nabla_z g), \tag{12}$$

with the Poisson tensor $\boldsymbol{B}(z)$ given by

$$\boldsymbol{B}(z) = \begin{bmatrix} \hat{\boldsymbol{\Pi}} & \hat{\boldsymbol{\gamma}} & \hat{\boldsymbol{R}} & \hat{\boldsymbol{P}} \\ \hat{\boldsymbol{\gamma}} & 0 & 0 & 0 \\ \hat{\boldsymbol{R}} & 0 & 0 & \boldsymbol{E} \\ \hat{\boldsymbol{P}} & 0 & -\boldsymbol{E} & 0 \end{bmatrix}, \tag{13}$$

where $\boldsymbol{E}$ is the identity matrix. This Poisson tensor has two independent Casimir functions. One is a geometric integral $C_1(z) = \boldsymbol{\gamma}^T \boldsymbol{\gamma}/2 \equiv 1/2$, and the other one is $C_2(z) = \boldsymbol{\gamma}^T \left(\boldsymbol{\Pi} + \hat{\boldsymbol{R}}\boldsymbol{P}\right)$, the third component of the angular momentum with respect to origin $O$ expressed in the inertial frame $S$. $C_2(z)$ is the conservative quantity produced by the symmetry of the system.

The ten-dimensional invariant manifold or symplectic leaf of the system is defined in $\mathbb{R}^{12}$ by Casimir functions

$$\Sigma = \left\{ \left(\boldsymbol{\Pi}^T, \boldsymbol{\gamma}^T, \boldsymbol{R}^T, \boldsymbol{P}^T\right)^T \in \mathbb{R}^{12} \mid \boldsymbol{\gamma}^T\boldsymbol{\gamma} = 1, \boldsymbol{\gamma}^T\left(\boldsymbol{\Pi} + \hat{\boldsymbol{R}}\boldsymbol{P}\right) = \text{constant} \right\}, \tag{14}$$

which is actually the reduced phase space $T^*(Q/S^1)$ of the symplectic reduction. The restriction of the Poisson bracket $\{\cdot, \cdot\}_{\mathbb{R}^{12}}(z)$ to $\Sigma$ defines the symplectic structure on this symplectic leaf.

The equations of motion of the system can be written in the Hamiltonian form

$$\dot{z} = \{z, H(z)\}_{\mathbb{R}^{12}}(z) = \boldsymbol{B}(z)\nabla_z H(z). \tag{15}$$

With the Hamiltonian $H(z)$ given by Eq. (7), the explicit equations of motion are given by

$$\begin{aligned} \dot{\boldsymbol{\Pi}} &= \boldsymbol{\Pi} \times \boldsymbol{I}^{-1}\boldsymbol{\Pi} + \boldsymbol{R} \times \frac{\partial V(\boldsymbol{\gamma}, \boldsymbol{R})}{\partial \boldsymbol{R}} + \boldsymbol{\gamma} \times \frac{\partial V(\boldsymbol{\gamma}, \boldsymbol{R})}{\partial \boldsymbol{\gamma}}, \\ \dot{\boldsymbol{\gamma}} &= \boldsymbol{\gamma} \times \boldsymbol{I}^{-1}\boldsymbol{\Pi}, \\ \dot{\boldsymbol{R}} &= \boldsymbol{R} \times \boldsymbol{I}^{-1}\boldsymbol{\Pi} + \frac{\boldsymbol{P}}{m}, \\ \dot{\boldsymbol{P}} &= \boldsymbol{P} \times \boldsymbol{I}^{-1}\boldsymbol{\Pi} - \frac{\partial V(\boldsymbol{\gamma}, \boldsymbol{R})}{\partial \boldsymbol{R}}. \end{aligned} \tag{16}$$

Based on the equations of motion Eq. (16), we have obtained a classical type of relative equilibria of the rigid body under the second-order gravitational potential in Reference [23]. At this type of relative equilibria, the orbit of the mass center of the rigid body is a circle in the equatorial plane of body $P$ with its center coinciding with origin $O$. The rigid body rotates uniformly around one of its principal axes that is parallel to $e_3$ in the inertial frame $S$ in angular velocity that is equal to the orbital angular velocity $\Omega_e$. The position vector $\boldsymbol{R}_e$ and the linear momentum $\boldsymbol{P}_e$ are parallel to another two principal axes of the rigid body.

When the position vector $R_e$ is parallel to the principal axes of the rigid body $i, j, k$, the norm of the orbital angular velocity $\Omega_e$ is given by the following three equations respectively:

$$\Omega_e = \left( \frac{GM_1}{R_e^3} + \frac{3GM_1}{2R_e^5} \left[ -2\frac{I_{xx}}{m} + \frac{I_{yy}}{m} + \frac{I_{zz}}{m} + \varepsilon \right] \right)^{1/2}, \tag{17}$$

$$\Omega_e = \left( \frac{GM_1}{R_e^3} + \frac{3GM_1}{2R_e^5} \left[ \frac{I_{xx}}{m} - 2\frac{I_{yy}}{m} + \frac{I_{zz}}{m} + \varepsilon \right] \right)^{1/2}, \tag{18}$$

$$\Omega_e = \left( \frac{GM_1}{R_e^3} + \frac{3GM_1}{2R_e^5} \left[ \frac{I_{xx}}{m} + \frac{I_{yy}}{m} - 2\frac{I_{zz}}{m} + \varepsilon \right] \right)^{1/2}. \tag{19}$$

The norm of the linear momentum $P_e$ is given by:

$$P_e = mR_e\Omega_e. \tag{20}$$

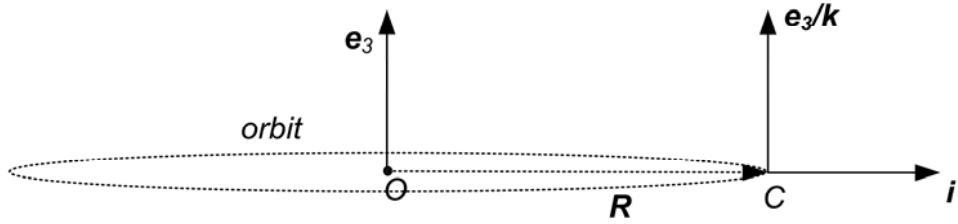

**Figure 2. One of the classical type of relative equilibria.**

With a given value of $R_e$, there are 24 relative equilibria belonging to this classical type in total. Without of loss of generality, we will choose one of the relative equilibria as shown by Figure 2 for stability conditions

$$\boldsymbol{\Pi}_e = [0, 0, \Omega_e I_{zz}]^T, \boldsymbol{\gamma}_e = [0, 0, 1]^T, \boldsymbol{R}_e = [R_e \ \ 0 \ \ 0]^T, \boldsymbol{P}_e = [0 \ \ mR_e\Omega_e \ \ 0]^T, \boldsymbol{\Omega}_e = [0 \ \ 0 \ \ \Omega_e]^T. \tag{21}$$

Other relative equilibria can be converted into this equilibrium by changing the arrangement of the axes of the reference frame $S_b$.

## LINEAR STABILITY OF THE RELATIVE EQUILIBRIA

In this section, we will investigate the linear stability of the relative equilibria through the linear system matrix using the method provided by the geometric mechanics[24][28].

### Conditions of Linear Stability

The linear stability of the relative equilibrium $z_e$ depends on the eigenvalues of the linear system matrix of the system at the relative equilibrium. According to Reference [24], the linear system matrix $D(z_e)$ of the non-canonical Hamiltonian system at the relative equilibrium $z_e$ can be calculated through the multiplication of the Poisson tensor and the Hessian of the variational Lagrangian without performing linearization as follows:

$$D(z_e) = B(z_e)\nabla^2 F(z_e). \tag{22}$$

Here the variational Lagrangian $F(z)$ is defined as

$$F(z) = H(z) - \sum_{i=1}^{2} \mu_i C_i(z). \tag{23}$$

According to Reference [24], the relative equilibrium of the rigid body in the problem corresponds to the stationary point of the Hamiltonian constrained by the Casimir functions. The stationary points can be determined by the first variation condition of the variational Lagrangian $\nabla F(z_e) = \mathbf{0}$. By using the formulations of the Hamiltonian and Casimir functions, the equilibrium conditions are obtained as:

$$\begin{aligned} & \mathbf{I}^{-1} \mathbf{\Pi}_e - \mu_2 \boldsymbol{\gamma}_e = \mathbf{0}, \\ & \frac{3GM_1 \varepsilon m}{R_e^3} (\boldsymbol{\gamma}_e \cdot \overline{\mathbf{R}}_e) \overline{\mathbf{R}}_e - \mu_1 \boldsymbol{\gamma}_e - \mu_2 \left( \mathbf{\Pi}_e + \hat{\mathbf{R}}_e \mathbf{P}_e \right) = \mathbf{0}, \\ & -\mu_2 \hat{\mathbf{P}}_e \boldsymbol{\gamma}_e + \left. \frac{\partial V}{\partial \mathbf{R}} \right|_e = \mathbf{0}, \\ & -\mu_2 \hat{\boldsymbol{\gamma}}_e \mathbf{R}_e + \frac{\mathbf{P}_e}{m} = \mathbf{0}. \end{aligned} \tag{24}$$

As we expected, the relative equilibrium in Eq. (21) obtained based on the equations of motion is a solution of the equilibrium conditions Eq. (24), with the parameters $\mu_1$ and $\mu_2$ given by

$$\mu_1 = -\Omega_e^2 \left( I_{zz} + m R_e^2 \right), \; \mu_2 = \Omega_e. \tag{25}$$

By using the formulation of the second-order gravitational potential Eq. (8), the Hessian of the variational Lagrangian $\nabla^2 F(z)$ is calculated as:

$$\nabla^2 F(z) = \begin{bmatrix} \mathbf{I}^{-1} & -\mu_2 \mathbf{I}_{3\times 3} & \mathbf{0} & \mathbf{0} \\ -\mu_2 \mathbf{I}_{3\times 3} & \frac{3GM_1 \varepsilon m}{R^5} \mathbf{R}\mathbf{R}^T - \mu_1 \mathbf{I}_{3\times 3} & \left( \frac{\partial^2 V}{\partial \boldsymbol{\gamma} \partial \mathbf{R}} \right)^T + \mu_2 \hat{\mathbf{P}} & -\mu_2 \hat{\mathbf{R}} \\ \mathbf{0} & \frac{\partial^2 V}{\partial \boldsymbol{\gamma} \partial \mathbf{R}} - \mu_2 \hat{\mathbf{P}} & \frac{\partial^2 V}{\partial \mathbf{R}^2} & \mu_2 \hat{\boldsymbol{\gamma}} \\ \mathbf{0} & \mu_2 \hat{\mathbf{R}} & -\mu_2 \hat{\boldsymbol{\gamma}} & \frac{1}{m} \mathbf{I}_{3\times 3} \end{bmatrix}. \tag{26}$$

The second-order partial derivates of the gravitational potential in Eq. (26) are obtained as follows:

$$\frac{\partial^2 V}{\partial \boldsymbol{\gamma} \partial \mathbf{R}} = \frac{3GM_1 \varepsilon m}{R^4} \left[ (\boldsymbol{\gamma} \cdot \overline{\mathbf{R}}) \mathbf{I}_{3\times 3} + \boldsymbol{\gamma} \overline{\mathbf{R}}^T - 5(\boldsymbol{\gamma} \cdot \overline{\mathbf{R}}) \overline{\mathbf{R}} \overline{\mathbf{R}}^T \right], \tag{27}$$

$$\begin{aligned} \frac{\partial^2 V}{\partial \mathbf{R}^2} &= \frac{GM_1 m}{R^3} \left( \mathbf{I}_{3\times 3} - 3 \overline{\mathbf{R}} \overline{\mathbf{R}}^T \right) \\ &+ \frac{3GM_1}{2R^5} \left\{ 5 \overline{\mathbf{R}}^T \mathbf{I} \overline{\mathbf{R}} - tr(\mathbf{I}) - \varepsilon m \left( 1 - 5(\boldsymbol{\gamma} \cdot \overline{\mathbf{R}})^2 \right) \right\} \left\{ 7 \overline{\mathbf{R}} \overline{\mathbf{R}}^T - \mathbf{I}_{3\times 3} \right\} \\ &+ \frac{3GM_1}{R^5} \left\{ \left[ tr(\mathbf{I}) + \varepsilon m \right] \overline{\mathbf{R}} \overline{\mathbf{R}}^T + \mathbf{I} + \varepsilon m \boldsymbol{\gamma} \boldsymbol{\gamma}^T \right\} \\ &+ \frac{15GM_1}{R^5} \left\{ -\mathbf{I} \overline{\mathbf{R}} \overline{\mathbf{R}}^T - \overline{\mathbf{R}} \overline{\mathbf{R}}^T \mathbf{I} - \varepsilon m (\boldsymbol{\gamma} \cdot \overline{\mathbf{R}}) (\boldsymbol{\gamma} \overline{\mathbf{R}}^T + \overline{\mathbf{R}} \boldsymbol{\gamma}^T) \right\}. \end{aligned} \tag{28}$$

As described by Eqs. (17), (21) and (25), at the relative equilibrium $z_e$, we have $\boldsymbol{\Pi}_e = [0, 0, \Omega_e I_{zz}]^T$, $\boldsymbol{\gamma}_e = [0, 0, 1]^T$, $\boldsymbol{R}_e = [R_e, 0, 0]^T$, $\bar{\boldsymbol{R}}_e = [1, 0, 0]^T$, $\boldsymbol{P}_e = [0, mR_e\Omega_e, 0]^T$, $\boldsymbol{\Omega}_e = [0, 0, \Omega_e]^T$, $\mu_1 = -\Omega_e^2(I_{zz} + mR_e^2)$ and $\mu_2 = \Omega_e$. Then the Hessian of the variational Lagrangian $\nabla^2 F(z_e)$ at the relative equilibrium $z_e$ can be obtained as:

$$\nabla^2 F(z_e) = \begin{bmatrix} \boldsymbol{I}^{-1} & -\mu_2 \boldsymbol{I}_{3\times3} & \boldsymbol{0} & \boldsymbol{0} \\ -\mu_2 \boldsymbol{I}_{3\times3} & \dfrac{3GM_1\varepsilon m}{R_e^5}\boldsymbol{R}_e \boldsymbol{R}_e^T - \mu_1 \boldsymbol{I}_{3\times3} & \left(\dfrac{\partial^2 V}{\partial \boldsymbol{\gamma} \partial \boldsymbol{R}}\right)^T\bigg|_e + \mu_2 \hat{\boldsymbol{P}}_e & -\mu_2 \hat{\boldsymbol{R}}_e \\ \boldsymbol{0} & \dfrac{\partial^2 V}{\partial \boldsymbol{\gamma} \partial \boldsymbol{R}}\bigg|_e - \mu_2 \hat{\boldsymbol{P}}_e & \dfrac{\partial^2 V}{\partial \boldsymbol{R}^2}\bigg|_e & \mu_2 \hat{\boldsymbol{\gamma}}_e \\ \boldsymbol{0} & \mu_2 \hat{\boldsymbol{R}}_e & -\mu_2 \hat{\boldsymbol{\gamma}}_e & \dfrac{1}{m}\boldsymbol{I}_{3\times3} \end{bmatrix}. \quad (29)$$

The second-order partial derivates of the gravitational potential in Eq. (29) at the relative equilibrium $z_e$ are obtained through Eqs. (27)-(28) as follows:

$$\left.\frac{\partial^2 V}{\partial \boldsymbol{\gamma} \partial \boldsymbol{R}}\right|_e = \frac{3GM_1\varepsilon m}{R_e^4}\boldsymbol{\gamma}_e \boldsymbol{\alpha}_e^T, \quad (30)$$

$$\left.\frac{\partial^2 V}{\partial \boldsymbol{R}^2}\right|_e = \frac{GM_1 m}{R_e^3}\left(\boldsymbol{I}_{3\times3} - 3\boldsymbol{\alpha}_e \boldsymbol{\alpha}_e^T\right) + \frac{3GM_1}{2R_e^5}\left\{\begin{matrix}\left[15I_{xx} - 5tr(\boldsymbol{I}) - 5\varepsilon m\right]\boldsymbol{\alpha}_e \boldsymbol{\alpha}_e^T + 2\varepsilon m \boldsymbol{\gamma}_e \boldsymbol{\gamma}_e^T \\ +2\boldsymbol{I} - \left[5I_{xx} - tr(\boldsymbol{I}) - \varepsilon m\right]\boldsymbol{I}_{3\times3}\end{matrix}\right\}, \quad (31)$$

where $\boldsymbol{\alpha}_e$ is defined as $\boldsymbol{\alpha}_e = [1 \ 0 \ 0]^T$.

The Poisson tensor $\boldsymbol{B}(z_e)$ at the relative equilibrium $z_e$ can be obtained as:

$$\boldsymbol{B}(z_e) = \begin{bmatrix} \Omega_e I_{zz}\hat{\boldsymbol{\gamma}}_e & \hat{\boldsymbol{\gamma}}_e & R_e\hat{\boldsymbol{\alpha}}_e & mR_e\Omega_e\hat{\boldsymbol{\beta}}_e \\ \hat{\boldsymbol{\gamma}}_e & \boldsymbol{0} & \boldsymbol{0} & \boldsymbol{0} \\ R_e\hat{\boldsymbol{\alpha}}_e & \boldsymbol{0} & \boldsymbol{0} & \boldsymbol{E} \\ mR_e\Omega_e\hat{\boldsymbol{\beta}}_e & \boldsymbol{0} & -\boldsymbol{E} & \boldsymbol{0} \end{bmatrix}, \quad (32)$$

where $\boldsymbol{\beta}_e$ is defined as $\boldsymbol{\beta}_e = [0, 1, 0]^T$. In Eqs. (29)-(32), we have

$$\hat{\boldsymbol{\alpha}}_e = \begin{bmatrix} 0 & 0 & 0 \\ 0 & 0 & -1 \\ 0 & 1 & 0 \end{bmatrix}, \ \hat{\boldsymbol{\beta}}_e = \begin{bmatrix} 0 & 0 & 1 \\ 0 & 0 & 0 \\ -1 & 0 & 0 \end{bmatrix}, \ \hat{\boldsymbol{\gamma}}_e = \begin{bmatrix} 0 & -1 & 0 \\ 1 & 0 & 0 \\ 0 & 0 & 0 \end{bmatrix}, \ \boldsymbol{\alpha}_e \boldsymbol{\alpha}_e^T = \begin{bmatrix} 1 & 0 & 0 \\ 0 & 0 & 0 \\ 0 & 0 & 0 \end{bmatrix}, \quad (33)$$

$$\boldsymbol{\gamma}_e \boldsymbol{\gamma}_e^T = \begin{bmatrix} 0 & 0 & 0 \\ 0 & 0 & 0 \\ 0 & 0 & 1 \end{bmatrix}, \ \boldsymbol{\gamma}_e \boldsymbol{\alpha}_e^T = \begin{bmatrix} 0 & 0 & 0 \\ 0 & 0 & 0 \\ 1 & 0 & 0 \end{bmatrix}. \quad (34)$$

Then the linear system matrix $\boldsymbol{D}(z_e)$ of the non-canonical Hamiltonian system can be calculated through Eqs. (22), (29) and (32). Through some rearrangement and simplification, the linear system matrix $\boldsymbol{D}(z_e)$ can be written as follows:

$$D(z_e) =$$

$$\begin{bmatrix} \Omega_e \left( I_{zz} \hat{\gamma}_e I^{-1} - \hat{\gamma}_e \right) & 0 & \left\{ \dfrac{GM_1 m}{R_e^2} - mR_e \Omega_e^2 - \dfrac{3GM_1}{2R_e^4} \left[ 5I_{xx} - tr(I) - \varepsilon m \right] \right\} \hat{a}_e + \dfrac{3GM_1}{R_e^4} \hat{a}_e I & 0 \\ \hat{\gamma}_e I^{-1} & -\Omega_e \hat{\gamma}_e & 0 & 0 \\ R_e \hat{a}_e I^{-1} & 0 & -\Omega_e \hat{\gamma}_e & \dfrac{1}{m} \mathbf{I}_{3\times 3} \\ mR_e \Omega_e \hat{\beta}_e I^{-1} & -\left. \dfrac{\partial^2 V}{\partial \gamma \partial \mathbf{R}} \right|_e & -\left. \dfrac{\partial^2 V}{\partial \mathbf{R}^2} \right|_e & -\Omega_e \hat{\gamma}_e \end{bmatrix} \quad (35)$$

As stated above, the linear stability of the relative equilibrium $z_e$ depends on the eigenvalues of the linear system matrix of the system $D(z_e)$. The characteristic polynomial of the linear system matrix $D(z_e)$ can be calculated by

$$P(s) = \det\left[ s\mathbf{I}_{12\times 12} - D(z_e) \right]. \tag{36}$$

The eigenvalues of the linear system matrix $D(z_e)$ are roots of the characteristic equation of the linearized system, which is given by

$$\det\left[ s\mathbf{I}_{12\times 12} - D(z_e) \right] = 0. \tag{37}$$

Through Eqs. (35) and (37), with the help of *Matlab* and *Maple*, the characteristic equation can be obtained with the following form:

$$s^2(m^2 I_{zz} s^4 + A_2 s^2 + A_0)(mI_{xx} I_{yy} s^6 + B_4 s^4 + B_2 s^2 + B_0) = 0, \tag{38}$$

where the coefficients $A_2$, $A_0$, $B_4$, $B_2$ and $B_0$ are functions of the parameters of the system: $GM_1$, $\Omega_e$, $R_e$, $\varepsilon$, $m$, $I_{xx}$, $I_{yy}$ and $I_{zz}$. The formulations of the coefficients are given in the Appendix.

According to Reference [24], the non-canonical Hamiltonian systems have special properties with regard to both the form of the characteristic polynomial and the eigenvalues of the linear system matrix $D(z_e)$:

**Property 1.** *There are only even terms in the characteristic polynomial of the linear system matrix, and the eigenvalues are symmetrical with respect to both the real and imaginary axes.*

**Property 2.** *A zero eigenvalue exists for each linearly independent Casimir function.*

**Property 3.** *An additional pair of zero eigenvalues exists for each first integral, which is associated with a symmetry of the Hamiltonian by Noether's theorem.*

Notice that in our problem, there are two linearly independent Casimir functions, and the two zero eigenvalues correspond to the two Casimir functions $C_1(z)$ and $C_2(z)$. The remaining ten eigenvalues correspond to the motion constrained by the Casimir functions on the ten-dimensional invariant manifold $\Sigma$. We have carried out a Poisson reduction by means of the symmetry of the Hamiltonian, and expressed the dynamics on the reduced phase space. The additional pair of zero eigenvalues according to **Property 3** has been eliminated by the reduction process. Therefore, our results in Eq. (38) are consistent with these three properties stated above.

According to the characteristic equation in Eq. (38), the ten-dimensional linear system on the invariant manifold $\Sigma$ decouples into two entirely independent four- and six-dimensional subsystems under the second-order gravitational potential. It is worth our special attention that this is not the decoupling between the freedoms of the rotational motion and the orbital motion of the rigid body, since the orbit-rotation coupling is considered in our study. Actually, the four-dimensional subsystem and $s^2$ are the three freedoms of the orbital and rotational motions within the equatorial plane of the body P, and the other three freedoms, i.e. orbital and rotational motions outside the equatorial plane of the body P, constitute the six-dimensional subsystem.

The linear stability of the relative equilibria implies that there are no roots of the characteristic equation with positive real parts. According to **Property 1**, the linear stability requires all the roots to be purely imaginary, that is $s^2$ is real and negative. Therefore, in this case of a conservative system, we can only get the necessary conditions of the stability through the linear stability of the relative equilibria.

According to the theory of the roots of the second and third degree polynomial equation, that the $s^2$ in Eq. (38) is real and negative is equivalent to

$$\left(\frac{A_2}{m^2 I_{zz}}\right)^2 - \frac{4A_0}{m^2 I_{zz}} \geq 0, \ A_2 > 0, \ A_0 > 0; \tag{39}$$

$$\frac{1}{27}\left(-\frac{1}{3}\left(\frac{B_4}{mI_{xx}I_{yy}}\right)^2 + \frac{B_2}{mI_{xx}I_{yy}}\right)^3 + \frac{1}{4}\left(\frac{2}{27}\left(\frac{B_4}{mI_{xx}I_{yy}}\right)^3 - \frac{B_4 B_2}{3m^2 I_{xx}^2 I_{yy}^2} + \frac{B_0}{mI_{xx}I_{yy}}\right)^2 \leq 0, \tag{40}$$

$$B_4 > 0, \ B_2 > 0, \ B_0 > 0.$$

We have given the conditions of linear stability of the relative equilibria in Eqs. (39) and (40). Given a set of the parameters of the system, we can determine whether the relative equilibria are linear stable by using the stability criterion given above.

**Case Studies**

However, the expressions of coefficients $A_2$, $A_0$, $B_4$, $B_2$ and $B_0$ in terms of the parameters of the system are tedious, since there are large amount of parameters in the system and the considered problem is a high-dimensional system. It is difficult to get general conditions of linear stability through Eqs. (39) and (40) in terms of the parameters of the system.

We will consider an example planet P, which has the same mass and equatorial radius as the Earth, but has a different zonal harmonic $J_2$. That is $GM_1 = 3.986005 \times 10^{14} \text{ m}^3/\text{s}^2$ and $a_E = 6.37814 \times 10^6 \text{ m}$. Five different values of the zonal harmonic $J_2$ are considered

$$J_2 = 0.5, 0.2, 0, -0.18, -0.2. \tag{41}$$

The orbital angular velocity $\Omega_e$ is assumed to be equal to $1.163553 \times 10^{-3} \text{s}^{-1}$ with the orbital period equal to 1.5 hours. With the parameters of the system given above, the stability criterion in Eqs. (39) and (40) can be determined by three mass distribution parameters of the rigid body: $I_{xx}/m$, $\sigma_x$ and $\sigma_y$, where $\sigma_x$ and $\sigma_y$ are defined as

$$\sigma_x = \left(\frac{I_{zz} - I_{yy}}{I_{xx}}\right), \ \sigma_y = \left(\frac{I_{zz} - I_{xx}}{I_{yy}}\right). \tag{42}$$

The ratio $I_{xx}/m$ describes the characteristic dimension of the rigid body; the ratios $\sigma_x$ and $\sigma_y$ describe the shape of the rigid body to the second order. Three different values of the parameter $I_{xx}/m$ are considered as follows:

$$\frac{I_{xx}}{m} = 5\times 10^{3}, 5\times 10^{7}, 5\times 10^{11}, \tag{43}$$

which correspond to a rigid body with the characteristic dimension of order of 100m, 10km and 1000km respectively. In the case of each value of $I_{xx}/m$, the parameters $\sigma_x$ and $\sigma_y$ are considered in the following range

$$-1 \leq \sigma_x \leq 1, -1 \leq \sigma_y \leq 1, \tag{44}$$

which have covered all the possible mass distributions of the rigid body.

Given the mass distribution parameters of the rigid body, the orbital radius $R_e$ at the relative equilibrium can be calculated by Eq. (17). Then the stability criterion in Eqs. (39) and (40) can be calculated with all the parameters of the system known. The linear stability criterion in Eqs. (39) and (40) is calculated for a rigid body within the range of the parameters Eqs. (43) and (44) in the cases of different values of the zonal harmonic $J_2$. The points, which correspond to the mass distribution parameters guaranteeing the linear stability, are plotted on the $\sigma_y - \sigma_x$ plane in the 15 cases of different values of $I_{xx}/m$ and $J_2$ in Figures (3)-(17) respectively.

In our problem, the gravitational potential in Eq. (8) is truncated on the second order. According to the conclusions in Reference [29], only the central component of the gravity field of the planet P is considered in the gravity gradient torque, with the zonal harmonic $J_2$ neglected. That is to say, the attitude motion of the rigid body in our problem is actually the attitude dynamics on a circular orbit in a central gravity field from the point view of the traditional attitude dynamics with the orbit-rotation coupling neglected. To make comparisons with the traditional attitude dynamics, we also plot the classical linear attitude stability region of a rigid body on a circular orbit in a central gravity field in Figures (3)-(17), which is given by:

$$\begin{aligned}\sigma_y - \sigma_x &> 0,\\ 1 + 3\sigma_y + \sigma_x\sigma_y &> 4\sqrt{\sigma_x\sigma_y},\\ \sigma_x\sigma_y &> 0.\end{aligned} \tag{45}$$

The classical linear attitude stability region given by Eq. (45) is consisted of the Lagrange region I and the DeBra-Delp region II[30]. The Lagrange region is the isosceles right triangle region in the first quadrant of the $\sigma_y - \sigma_x$ plane below the straight line $\sigma_y - \sigma_x = 0$, and DeBra-Delp region is a small region in the third quadrant below the straight line $\sigma_y - \sigma_x = 0$. Notice that at the relative equilibrium in our paper, the orientations of the principal axes of the rigid body are different from those at the equilibrium attitude in Reference [30], and then the definitions of the parameters $\sigma_y$ and $\sigma_x$ in our paper are different from those in Reference [30] to make sure that the linear attitude stability region is the same with that in Reference [30].

| 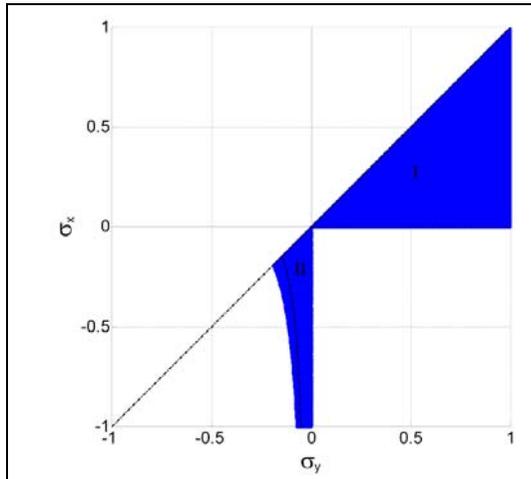 | 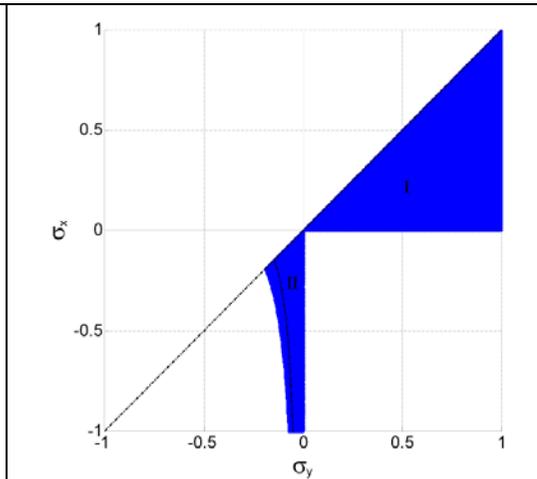 | 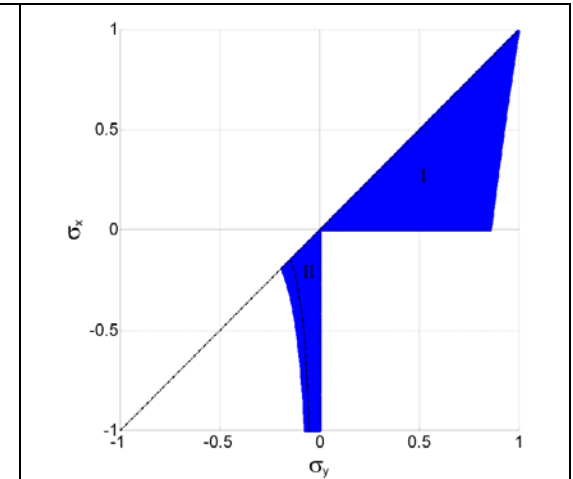 |
|---|---|---|
| **Figure 3. Linear stability region in the case of** $J_2 = 0.5$ **and** $I_{xx}/m = 5 \times 10^3$. | **Figure 4. Linear stability region in the case of** $J_2 = 0.5$ **and** $I_{xx}/m = 5 \times 10^7$. | **Figure 5. Linear stability region in the case of** $J_2 = 0.5$ **and** $I_{xx}/m = 5 \times 10^{11}$. |
| 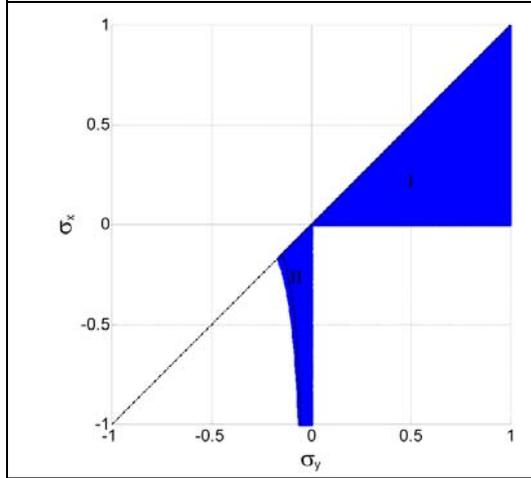 | 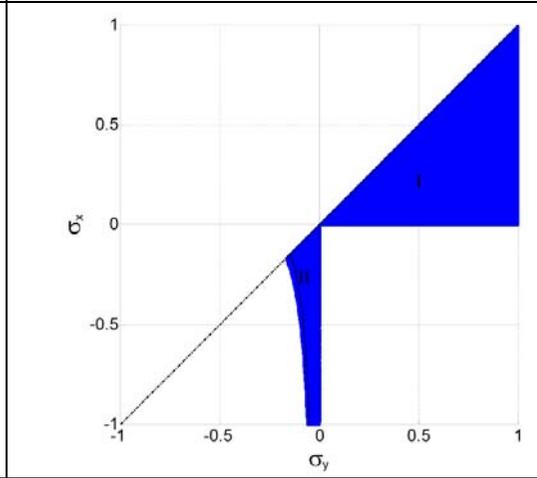 | 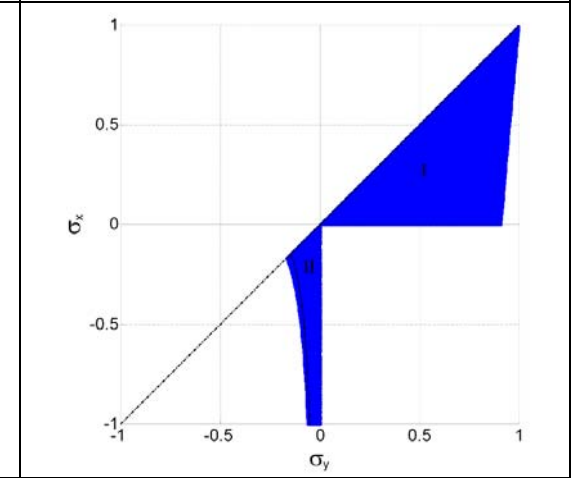 |
| **Figure 6. Linear stability region in the case of** $J_2 = 0.2$ **and** $I_{xx}/m = 5 \times 10^3$. | **Figure 7. Linear stability region in the case of** $J_2 = 0.2$ **and** $I_{xx}/m = 5 \times 10^7$. | **Figure 8. Linear stability region in the case of** $J_2 = 0.2$ **and** $I_{xx}/m = 5 \times 10^{11}$. |

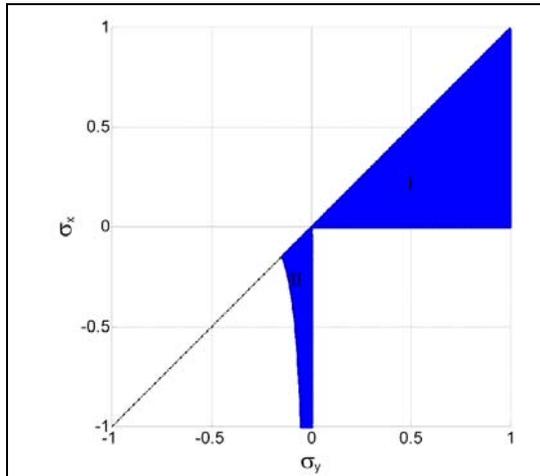
**Figure 9.** Linear stability region in the case of $J_2 = 0$ and $I_{xx}/m = 5 \times 10^3$.

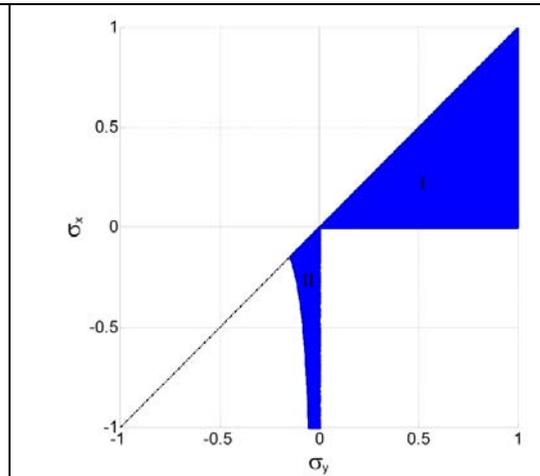
**Figure 10.** Linear stability region in the case of $J_2 = 0$ and $I_{xx}/m = 5 \times 10^7$.

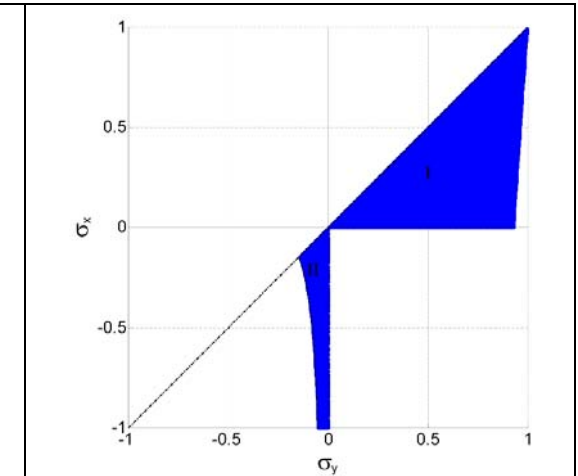
**Figure 11.** Linear stability region in the case of $J_2 = 0$ and $I_{xx}/m = 5 \times 10^{11}$.

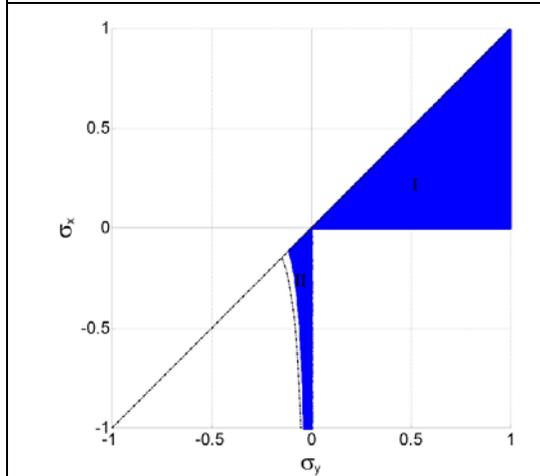
**Figure 12.** Linear stability region in the case of $J_2 = -0.18$ and $I_{xx}/m = 5 \times 10^3$.

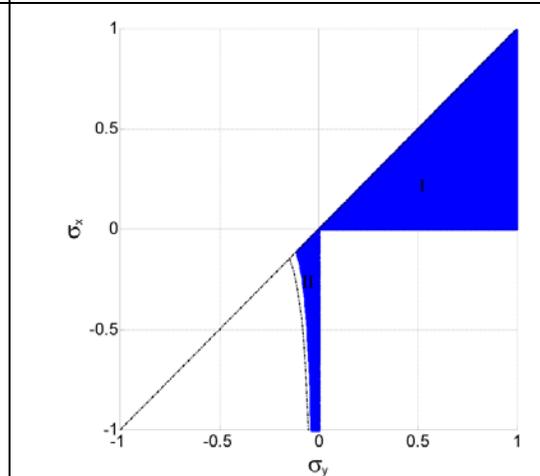
**Figure 13.** Linear stability region in the case of $J_2 = -0.18$ and $I_{xx}/m = 5 \times 10^7$.

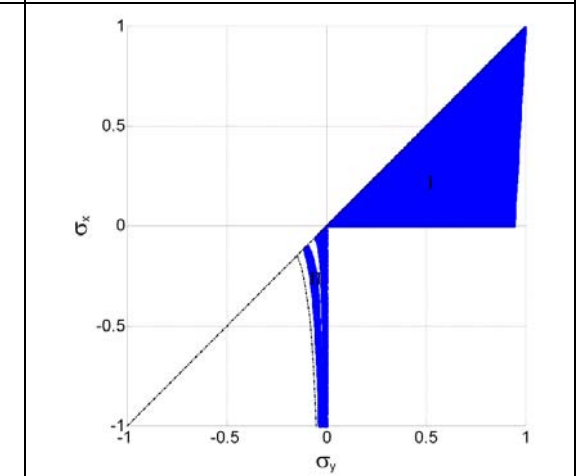
**Figure 14.** Linear stability region in the case of $J_2 = -0.18$ and $I_{xx}/m = 5 \times 10^{11}$.

| 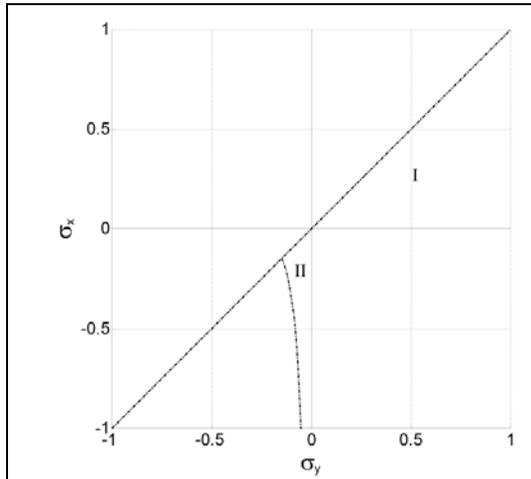 | 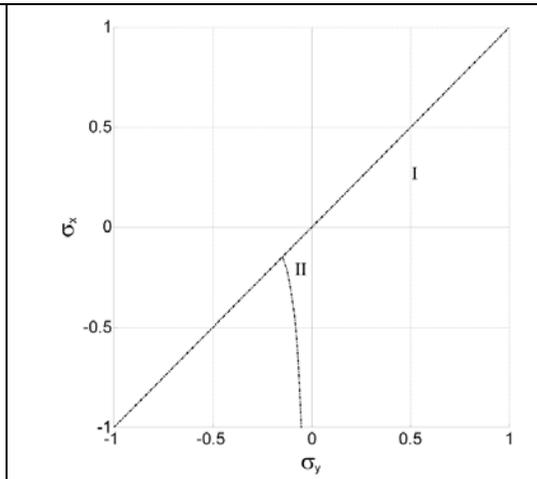 | 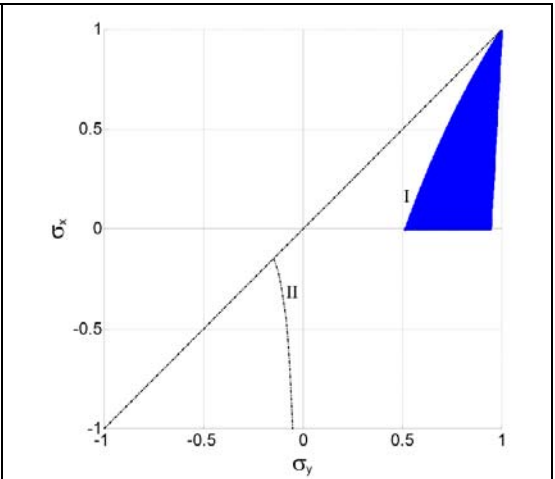 |
|---|---|---|
| **Figure 15.** Linear stability region in the case of $J_2 = -0.2$ and $I_{xx}/m = 5\times10^3$. | **Figure 16.** Linear stability region in the case of $J_2 = -0.2$ and $I_{xx}/m = 5\times10^7$. | **Figure 17.** Linear stability region in the case of $J_2 = -0.2$ and $I_{xx}/m = 5\times10^{11}$. |

**Some Discussions on the Linear Stability**

From Figures (3)-(17), we can easily achieve several conclusions as follows:

(a). Similar to the classical linear attitude stability region, which is consisted of the Lagrange region and the DeBra-Delp region, the linear stability region of the relative equilibrium of the rigid body in our problem is also consisted of two regions located in the first and third quadrant of the $\sigma_y - \sigma_x$ plane, which are the analogues of the Lagrange region and the DeBra-Delp region respectively. This is consistent with the conclusion by Teixidó Román[5] that for a rigid body in a central gravity field there is a linear stability region in the third quadrant of the $\sigma_y - \sigma_x$ plane, which is the analogue of the DeBra-Delp region. However, when the planet $P$ is very elongated with $J_2 = -0.2$, for a small rigid body there is no linear stability region; only in the case of a very large rigid body with $I_{xx}/m = 5 \times 10^{11}$, there is a linear stability region that is the analogue of the Lagrange region located in the first quadrant of the $\sigma_y - \sigma_x$ plane.

(b). For a given value of the zonal harmonic $J_2$ (except $J_2 = -0.2$), when the characteristic dimension of the rigid body is small, the characteristic dimension of the rigid body have no influence on the linear stability region, as shown by the linear stability region in the cases of $I_{xx}/m = 5 \times 10^3$ and $I_{xx}/m = 5 \times 10^7$. In these cases, the linear stability region in the first quadrant of the $\sigma_y - \sigma_x$ plane, the analogue of the Lagrange region, is actually the Lagrange region. When the characteristic dimension of the rigid body is large enough, such as $I_{xx}/m = 5 \times 10^{11}$, the linear stability region in the first quadrant of the $\sigma_y - \sigma_x$ plane, the analogue of the Lagrange region, is reduced by a triangle in the right part of the first quadrant of the $\sigma_y - \sigma_x$ plane, as shown by Figures (5), (8), (11) and (14). In the case of $J_2 = -0.18$, also the linear stability region in the third quadrant of the $\sigma_y - \sigma_x$ plane, the analogue of the DeBra-Delp region, is reduced by the large characteristic dimension of the rigid body, as shown by Figure (14).

(c). For a given value of the characteristic dimension of the rigid body, as the zonal harmonic $J_2$ increases from -0.18 to 0.5, the linear stability region in the third quadrant of the $\sigma_y - \sigma_x$ plane, the analogue of the DeBra-Delp region, expands in the direction of the boundary of the DeBra-Delp region, and cross the boundary of the DeBra-Delp region at $J_2 = 0$. For a small value of the characteristic dimension of the rigid body, such as $I_{xx}/m = 5 \times 10^3$ and $I_{xx}/m = 5 \times 10^7$, as the zonal harmonic $J_2$ increases from -0.18 to 0.5, the linear stability region in the first quadrant of the $\sigma_y - \sigma_x$ plane, the analogue of the Lagrange region, keeps equal to the Lagrange region. Whereas for a large value of the characteristic dimension of the rigid body $I_{xx}/m = 5 \times 10^{11}$, as the zonal harmonic $J_2$ increases from -0.18 to 0.5, the linear stability region in the first quadrant of the $\sigma_y - \sigma_x$ plane, the analogue of the Lagrange region, is shrunk by the influence of the zonal harmonic $J_2$.

**CONCLUSION**

For new high-precision applications in celestial mechanics and astrodynamics, we have generalized the classical $J_2$ problem to the motion of a rigid body in a $J_2$ gravity field. Based on our previous results on the relative equilibria, the linear stability of the classical type of relative equilibria of this generalized problem is investigated in the framework of geometric mechanics.

The conditions of linear stability of the relative equilibria are obtained based on the characteristic equation of the linear system matrix at the relative equilibria, which is given through the multiplication of the Poisson tensor and Hessian matrix of the variational Lagrangian. With the stability conditions, the linear stability of the relative equilibria is investigated in a wide range of the parameters of the gravity field and the rigid body by using the numerical method. The stability region is plotted on the plane of the mass distribution parameters of the rigid body in the cases of different values of the zonal harmonic $J_2$ and the characteristic dimension of the rigid body.

Similar to the classical attitude stability in a central gravity field, the linear stability region is consisted of two regions located in the first and third quadrant of the $\sigma_y - \sigma_x$ plane respectively, which are analogues of the Lagrange region and the DeBra-Delp region respectively. Both the zonal harmonic $J_2$ and the characteristic dimension of the rigid body have significant influences on the linear stability. When the characteristic dimension of the rigid body is small, the analogue of the Lagrange region in the first quadrant of the $\sigma_y - \sigma_x$ plane is actually the Lagrange region. When the characteristic dimension of the rigid body is large enough, the analogue of the Lagrange region is reduced by a triangle and this triangle expands as the zonal harmonic $J_2$ increases. For a given value of the characteristic dimension of the rigid body, as the zonal harmonic $J_2$ increases, the analogue of the DeBra-Delp region in the third quadrant of the $\sigma_y - \sigma_x$ plane expands in the direction of the boundary of the DeBra-Delp region, and cross the boundary of the DeBra-Delp region at $J_2 = 0$.

Our results on the stability of the relative equilibria are very useful for studies on the motion of many natural satellites in our solar system, whose motions are close to the relative equilibria.

**ACKNOWLEDGMENTS**

This work is supported by the Innovation Foundation of BUAA for PhD Graduates, the Graduate Innovation Practice Foundation of BUAA under Grant YCSJ-01-201306 and the Fundamental Research Funds for the Central Universities.

**APPENDIX: THE COEFFICIENTS IN CHARACTERISTIC EQUATION**

The explicit formulations of the coefficients in the characteristic equations Eq. (38) are given as follows:

$$A_2 = -\frac{m}{2R_e^5}\big(3I_{yy}I_{zz}\mu + 12R_e^2 mI_{xx}\mu + 9mI_{zz}\mu\varepsilon - R_e^2 mI_{zz}\mu + 9I_{zz}^2\mu - 3R_e^2 m^2\mu\varepsilon$$
$$-12I_{xx}\mu I_{zz} - 4R_e^5 m\Omega_e^2 I_{zz} + 2m^2 R_e^7\Omega_e^2 - 9R_e^2 mI_{yy}\mu - 2R_e^4 m^2\mu\big), \tag{A.1}$$

$$A_0 = -\frac{1}{2R_e^{10}}\big(-9\mu I_{yy} + 12I_{xx}\mu - 2m\mu R_e^2 + 2mR_e^5\Omega_e^2 - 3m\mu\varepsilon - 3\mu I_{zz}\big)*$$
$$\big(3m^2 R_e^7\Omega_e^2 - 2R_e^4 m^2\mu - 6R_e^2 m^2\mu\varepsilon - R_e^5 m\Omega_e^2 I_{zz} - 8R_e^2 mI_{zz}\mu - 6R_e^2 mI_{yy}\mu$$
$$+12R_e^2 mI_{xx}\mu - 6mI_{zz}\mu\varepsilon - 6I_{yy}I_{zz}\mu + 12I_{xx}\mu I_{zz} - 6I_{zz}^2\mu\big), \tag{A.2}$$

$$B_4 = -\frac{1}{2R_e^5}\big(-5R_e^2 I_{yy}mI_{xx}\mu - 3I_{yy}^2 I_{xx}\mu + 2R_e^5 mI_{xx}I_{zz}\Omega_e^2 + 12I_{yy}I_{xx}^2\mu - 2R_e^4 m^2 I_{xx}\mu$$
$$-3R_e^2 m^2 I_{xx}\mu\varepsilon - 4R_e^5 I_{yy}mI_{xx}\Omega_e^2 + 2R_e^7 m^2 I_{xx}\Omega_e^2 + 2R_e^5 I_{yy}m\Omega_e^2 I_{zz}$$
$$-9I_{yy}mI_{xx}\mu\varepsilon - 9I_{yy}I_{xx}I_{zz}\mu - 2R_e^5 m\Omega_e^2 I_{zz}^2 + 12R_e^2 mI_{xx}^2\mu - 9R_e^2 mI_{xx}I_{zz}\mu\big), \tag{A.3}$$

$$B_2 = \frac{1}{2R_e^8}\bigl(-8I_{yy}mR_e^5\Omega_e^2 I_{zz}\mu + 27I_{yy}R_e^3 I_{xx}I_{zz}\mu\Omega_e^2 + 2m^2R_e^7 I_{xx}\Omega_e^2\mu - 5mR_e^5 I_{xx}I_{zz}\mu\Omega_e^2$$
$$+3m^2R_e^5\Omega_e^2 I_{zz}\mu\varepsilon - 2m^2R_e^{10}\Omega_e^4 I_{zz} + 6R_e^2 m^2 I_{xx}\mu^2\varepsilon + 19I_{yy}mR_e^5 I_{xx}\Omega_e^2\mu + 27mI_{xx}\mu^2\varepsilon I_{zz}$$
$$+11mR_e^5\Omega_e^2 I_{zz}^2\mu + 2m^2R_e^7\Omega_e^2 I_{zz}\mu - 3I_{yy}^2 mR_e^5\Omega_e^2\mu - 2I_{yy}m^2 R_e^7\Omega_e^2\mu - 2m^2 R_e^{10} I_{xx}\Omega_e^4$$
$$+2I_{yy}m^2 R_e^{10}\Omega_e^4 + 9R_e^3\Omega_e^2 I_{zz}^3\mu - 36mI_{xx}^2\mu^2\varepsilon + 9\mu^2 m^2\varepsilon^2 I_{xx} + 2mR_e^8\Omega_e^4 I_{zz}^2$$
$$-6I_{yy}R_e^3\Omega_e^2 I_{zz}^2\mu - 9I_{yy}mR_e^3\Omega_e^2 I_{zz}\mu\varepsilon - 9mR_e^3 I_{xx}\Omega_e^2 I_{zz}\mu\varepsilon - 21R_e^3 I_{xx}I_{zz}^2\mu\Omega_e^2$$
$$-2mR_e^8 I_{xx}\Omega_e^4 I_{zz} - 3I_{yy}m^2 R_e^5\Omega_e^2\mu\varepsilon - 3m^2 R_e^5 I_{xx}\Omega_e^2\mu\varepsilon - 24I_{yy}R_e^3 I_{xx}^2\Omega_e^2\mu$$
$$-3I_{yy}^2 R_e^3\Omega_e^2 I_{zz}\mu - 12mR_e^5 I_{xx}^2\Omega_e^2\mu + 6I_{yy}^2 R_e^3 I_{xx}\Omega_e^2\mu + 9mR_e^3\Omega_e^2 I_{zz}^2\mu\varepsilon - 2I_{yy}mR_e^8\Omega_e^4 I_{zz}$$
$$+2I_{yy}mR_e^8 I_{xx}\Omega_e^4 + 18I_{yy}mR_e^3 I_{xx}\Omega_e^2\mu\varepsilon + 9I_{yy}mI_{xx}\mu^2\varepsilon + 12R_e^3 I_{xx}^2 I_{zz}\mu\Omega_e^2\bigr), \quad \text{(A.4)}$$

$$B_0 = -\frac{\Omega_e^2}{2R_e^8}(I_{zz} - I_{yy})\bigl(-9R_e^3 I_{zz}^2\mu\Omega_e^2 - 11I_{zz}mR_e^5\Omega_e^2\mu - 3I_{yy}R_e^3 I_{zz}\mu\Omega_e^2 + 21I_{zz}R_e^3 I_{xx}\mu\Omega_e^2$$
$$-9I_{zz}mR_e^3\Omega_e^2\mu\varepsilon - 27I_{zz}m\mu^2\varepsilon + 14mR_e^5 I_{xx}\mu\Omega_e^2 - 9I_{yy}m\mu^2\varepsilon + 36mI_{xx}\mu^2\varepsilon - 6R_e^2 m^2\mu^2\varepsilon$$
$$-2m^2 R_e^7\Omega_e^2\mu - 3I_{yy}mR_e^5\Omega_e^2\mu - 12R_e^3 I_{xx}^2\mu\Omega_e^2 + 3I_{yy}R_e^3 I_{xx}\mu\Omega_e^2 - 9\mu^2 m^2\varepsilon^2$$
$$+9mR_e^3 I_{xx}\Omega_e^2\mu\varepsilon + 3m^2 R_e^5\Omega_e^2\mu\varepsilon + 2m^2 R_e^{10}\Omega_e^4\bigr), \quad \text{(A.5)}$$